\documentclass[preprint,english,preprintnumbers,amsmath,amssymb,floatfix]{revtex4}
\usepackage[T1]{fontenc}
\usepackage[latin9]{inputenc}
\usepackage{color}
\usepackage{array}
\usepackage{amstext}
\usepackage{graphicx}
\usepackage{esint}

\makeatletter

\@ifundefined{textcolor}{}
{%
 \definecolor{BLACK}{gray}{0}
 \definecolor{WHITE}{gray}{1}
 \definecolor{RED}{rgb}{1,0,0}
 \definecolor{GREEN}{rgb}{0,1,0}
 \definecolor{BLUE}{rgb}{0,0,1}
 \definecolor{CYAN}{cmyk}{1,0,0,0}
 \definecolor{MAGENTA}{cmyk}{0,1,0,0}
 \definecolor{YELLOW}{cmyk}{0,0,1,0}
 }

\@ifundefined{definecolor}
 {\usepackage{color}}{}
\@ifundefined{definecolor}
 {\usepackage{color}}{}
\makeatother

\makeatother

\usepackage{babel}

\begin{document}
%
\newcommand{\MSbar}{\ensuremath{\overline{\text{MS}}\ }}

\title{The effect of LHC ATLAS jet production cross sections data at $ \sqrt{s} = 7$~TeV on the proton PDFs up to N3LO}

\author{A.~Vafaee}
\email[]{vafaee.phy@gmail.com}
\affiliation{Department of Physics, Ferdowsi University of Mashhad, P.O.Box 1436, Mashhad, Iran}
\affiliation{Iran's National Elites Foundation, P. O. Box 14578-93111, Tehran, Iran}
\author{K.~Javidan}
\email[]{javidan@ferdowsi.um.ac.ir}
\affiliation{Department of Physics, Ferdowsi University of Mashhad, P.O.Box 1436, Mashhad, Iran}
\author{A.~B.~Shokouhi}
\email[]{shokouhi.phy@gmail.com}
\affiliation{Independent researcher, P. O. Box 11155-811, Tehran, Iran}

\date{\today}

\begin{abstract}
The effect of full $7$ sets of LHC ATLAS jet cross sections data at $ \sqrt{s} = 7$~TeV on the proton parton distribution functions (PDFs) up to next-to-next-to-next-to-leading order (NNNLO or N3LO) corrections is investigated for the first time. Phenomenologically, the proton central PDFs in this perturbative Quantum Chromo Dynamics (pQCD) analysis are defined based on the full seven data sets from HERA I and II combined. It is shown that, the LHC ATLAS jet cross sections data at $ \sqrt{s} = 7$~TeV on the HERA I and II combined data reduces the error band of proton PDFs. Particularly, the uncertainties of the gluon $xg(x,Q^2)$ and charm $xc(x,Q^2)$ distributions decrease dramatically. Adding the LHC ATLAS jet cross sections data at $ \sqrt{s} = 7$~TeV on the central proton PDFs improves the quality of the fit up to $\sim 1.53$~\%, $\sim 2.72$~\% and $\sim 2.80$~\% corresponding to next-to-leading order (NLO), next-to-next-to leading order (N2LO) and N3LO, respectively.       
  
\end{abstract}


\maketitle

\section{\label{introduction}Introduction}
In the quark-parton model, the proton is found to be a complex dynamical system composed of gluons, quarks and antiquarks, which are continuously interact with each other within in the pQCD framework theory. The structure functions of proton in the quark-parton model can be written as a convolution between hard scattering coefficients and PDFs, which are calculable and non-calculable parts of QCD theory, respectively. The proton PDFs as non-calculable parts of QCD theory are parametrized at an initial scale of $Q_0^2$ and determined by fitting to the experimental data. More information about quark-parton model can be found in Refs.~\cite{Bjorken:1969ja,GellMann:1964nj}. 

The quark-parton model for deep inelastic ${e^\pm}p$ scattering is formulated in a special frame where the proton has very high energy $E \gg m_p$ referred to as the infinite momentum frame. Accordingly, in the infinite momentum frame both the mass of the proton $m_p$ and any component of the momentum of the struck quark $p_q$ transverse to the direction of motion of the proton can be neglected. More details about infinite momentum can be found in Refs.~\cite{Aznaurian:1982qc,Leutwyler:1977vy}.  

The proton PDFs reflect the underlying internal structure of the proton and at the present time they cannot be determined from the first principles of pQCD theory. This is because the strong coupling constant of the QCD theory has a large value $\alpha_s \sim \cal O$($1$) and accordingly the pQCD theory is not suitable to determine the proton PDFs from the first principles. More information about the proton PDFs and proton structure functions can be found in Refs.~\cite{Lipatov:1974qm,Abramowicz:2015mha}.

In this N3LO QCD analysis to extract the proton PDFs by fitting to experimental data, we firstly parameterize the proton parton distributions at the starting scale of $Q_0^2$ based on the standard LHAPDF functional form style and then we evolve the parametrized PDFs using the DGLAP collinear evolution equations. 

To explore the internal structure of the proton and its quark-gluon dynamics as described within the pQCD framework, we use full seven set of HERA run I and II combined data from deep inelastic ${e^\pm}p$ scattering for neutral current (NC) and charged current (CC) reactions as our central data sets~\cite{Abramowicz:2015mha}. Generally, in order to obtain more direct information about the quark-gluon dynamics of the proton, combining data sets from different NC and CC reactions play a central role. 

As previously mentioned, we first parametrize the proton PDFs based on the HERAPDF standard functional form and then extract the proton central PDFs by fitting to HERA run I and II combined data using DGLAP collinear evolution equations at the next-to-leading order corrections. Then to obtain a direct constrain on the proton PDFs, particularly on the gluon and charm distributions, we add the full seven set of LHC ATLAS jet cross sections data at $ \sqrt{s} = 7$~TeV~\cite{Aad:2011fc} on the proton central PDFs up to N3LO corrections.

Within the quark-parton model at the pQCD level, we may obtain much more direct information about quark and gluon content of the proton by global fit of available experimental data from different NC and CC reactions at the DIS processes. The effect of LHC jet data on MSTW and MMHT PDFs at NLO and NNLO using reweighting techniques have been published in Refs.~\citep{Watt:2013oha,Harland-Lang:2017ytb}.

 In this article, we investigate the role and influence of $7$ sets of LHC ATLAS jet cross sections data at $ \sqrt{s} = 7$~TeV on the HERAPDF style PDFs up to N3LO by full fit procedure to estimate the impact of these new data on the proton PDFs.

 The outline of this  paper is as follows: In Sec.~(\ref{dis}) we describe the theoretical framework of deep inelastic $e^{\pm}p$ scattering  for NC and CC reactions. We introduce the theoretical framework for LHC ATLAS jet production cross sections in Sec.~(\ref{jet}). In Sec.~(\ref{fit}) we discuss about our QCD fit methodology and extract the proton PDFs. The QCD set-up and minimization procedure have been discussed in Sec.~(\ref{qcdsetup}). In Sec.~(\ref{res}) we present our QCD analysis results. We conclude with a summery and discussion in Sec.~(\ref{sum}).

\section{\label{dis}Theoretical framework for DIS of $e^{\pm}p$ scattering}

A combination of measurements of all inclusive deep inelastic $e^{\pm}p$ cross sections have been recently published by HERA run I and II combined data for NC and CC reactions at zero beam polarization~\cite{Abramowicz:2015mha}. A summary of full seven data sets of HERA I and II combined data is presented in Table~\ref{tab:hdata} where NDP refers to the number of data points.

The central proton PDFs of this pQCD analysis have been extracted by fitting to the data sets in Table~\ref{tab:hdata}.

\begin{table}[h]
\begin{center}
\begin{tabular}{|l|c|c|c|c|c|c|c|c|}
\hline
\hline
 {Experiment} & {NDP} & {$E_p$} & {$E_e$} & {${\cal L}$} & {${e^\pm}$} & $\sqrt{s}$ & {Current} & {Ref.} \\ \hline 
  HERA I+II CC $e^{+}p$ & 42 & $920$~GeV & $27.5$~GeV & $\sim 1$~pb$^{-1}$ & $e^+$ & $318$~GeV & CC & \cite{Abramowicz:2015mha} \\ \hline 
  HERA I+II CC $e^{-}p$ & 39 & $920$~GeV & $27.5$~GeV & $\sim 1$~pb$^{-1}$ & $e^-$ & $318$~GeV & CC & \cite{Abramowicz:2015mha} \\ \hline 
  HERA I+II NC $e^{-}p$ & 159 & $920$~GeV & $27.5$~GeV & $\sim 1$~pb$^{-1}$ & $e^-$ & $318$~GeV & NC & \cite{Abramowicz:2015mha} \\ \hline 
  HERA I+II NC $e^{+}p$ & 210 & $460$~GeV & $27.5$~GeV & $\sim 1$~pb$^{-1}$ & $e^+$ & $225$~GeV & NC & \cite{Abramowicz:2015mha}\\ \hline 
  HERA I+II NC $e^{+}p$ & 260 & $575$~GeV & $27.5$~GeV & $\sim 1$~pb$^{-1}$ & $e^+$ & $251$~GeV & NC & \cite{Abramowicz:2015mha} \\ \hline
  HERA I+II NC $e^{+}p$ & 112 & $820$~GeV & $27.5$~GeV & $1$~pb$^{-1}$ & $e^+$ & $318$~GeV & NC & \cite{Abramowicz:2015mha} \\ \hline 
  HERA I+II NC $e^{+}p$ & 485 & $920$~GeV & $27.5$~GeV & $\sim 1$~pb$^{-1}$ & $e^+$ & $318$~GeV & NC & \cite{Abramowicz:2015mha} \\ \hline \hline
  \end{tabular}
\vspace{-0.0cm}
\caption{\label{tab:hdata}{A summary of full seven data sets of HERA I and II combined.}}
\vspace{-0.4cm}
\end{center}
\end{table}   

The reduced NC and CC deep inelastic of $e^{\pm}p$ scattering cross sections are given in terms of a linear combination of generalized proton structure functions as follows:

 \begin{eqnarray}
 \label{eq:nc}
\sigma^{\pm}_{red,NC}(x,Q^2)&=&   \frac{d^2\sigma^{e^{\pm}p}_{NC}}{dxdQ^2}\cdot \frac{Q^4 x}{2\pi \alpha^2 (1 + (1-y)^2)}  \\                                          
    &=&  \tilde{F_2}(x,Q^2) \mp \frac{(1 - (1-y)^2)}{(1 + (1-y)^2)} x\tilde{F_3}(x,Q^2) -\frac{y^2}{(1 + (1-y)^2)} \tilde{F_{\rm L}}(x,Q^2)~, \nonumber
\end{eqnarray}

and 

\begin{eqnarray}
\label{eq:rcc}
\sigma^{\pm}_{red,CC}(x,Q^2)&=&\frac{d^2\sigma^{e^{\pm}p}_{CC}}{dxdQ^2}\cdot\frac{2\pi x}{G^2_F} \left[\frac{M^2_W+Q^2}{M^2_W}\right]^2  \\
  &=&\frac{(1 + (1-y)^2)}{2}  W^{\pm}_2(x,Q^2) - \frac{(1 - (1-y)^2)}{2}x  W^{\pm}_3(x,Q^2) - \frac{y^2}{2} W^{\pm}_L(x,Q^2)~, \nonumber
\end{eqnarray}

where $\tilde{F_2}(x,Q^2)$, $x \tilde{F_3}(x,Q^2)$ and $\tilde{F_L}(x,Q^2)$ are the generalized proton structure functions corresponding to NC reactions and $W^{\pm}_2(x,Q^2)$, $x  W^{\pm}_3(x,Q^2)$ and $W^{\pm}_L(x,Q^2)$ are the generalized proton structure functions corresponding to CC reactions. More details can be found in Ref.~\citep{Vafaee:2017nze}.

The double differential cross sections for NC and CC reactions, $\frac{d^2 \sigma_{\rm NC}}{dx dy}$ and $\frac{d^2 \sigma_{\rm CC}}{dx dy}$, which we use in this pQCD analysis are as follows:
\begin{eqnarray}
\label{eq:nc}
\frac{d^2 \sigma_{\rm NC}}{dx dy} &=& \frac{G_F^2}{16 \pi} \left[ \frac{M_Z^2}{Q^2 +M_Z^2}\right]^2
\left\{Y_+ W_2^{\rm NC}(x,Q^2) \pm Y_- xW_3^{\rm NC}(x,Q^2) - y^2 W_L^{\rm NC}(x,Q^2)\right\}\, ,
\\
\label{eq:cc}
\frac{d^2 \sigma_{\rm CC}}{dx dy} &=& \frac{G_F^2}{4 \pi} \left[ \frac{M_W^2}{Q^2 + M_W^2}\right]^2
\left\{Y_+ W_2^{\rm CC}(x,Q^2) \pm Y_- xW_3^{\rm CC}(x,Q^2) - y^2 W_L^{\rm CC}(x,Q^2)\right\}\, ,
\end{eqnarray}
where as before $G_F$ denotes the Fermi constant and $Y_\pm = 1 \pm (1-y)^2$.

It should be noted that in this pQCD analysis we perform four different fits entitled: HNLO, HANLO, HAN2LO and HAN3LO so that in throughout of this article the words HNLO, HANLO, HAN2LO and HAN3LO refer as follows:
\begin{itemize}
\item {\bf HNLO:} HERA I and II combined data at the NLO corrections.
\item {\bf HANLO:} HERA I and II combined data plus LHC ATLAS inclusive jet production cross sections data sets at the NLO corrections.
\item {\bf HAN2LO:} HERA I and II combined data plus LHC ATLAS inclusive jet production cross sections data sets at the N2LO corrections.
\item {\bf HAN3LO:} HERA I and II combined data plus LHC ATLAS inclusive jet production cross sections data sets at the N3LO corrections.
\end{itemize}

\section{\label{jet}Theoretical framework for the LHC ATLAS jet production cross sections}

The LHC ATLAS inclusive jet production cross sections data have been provided in proton-proton collision with a total integrated luminosity of ${\cal L} = 37$~pb$^{-1}$ at a centre-of-mass energy of $\sqrt{s} = 7$~TeV. LHC ATLAS inclusive jet production cross sections data were measured using jets clustered with two radius parameters $R = 0.4$ and $R = 0.6$~\cite{Aad:2011fc}. Phenomenologically, we cannot use both $R = 0.4$ and $R = 0.6$ at the same time. Accordingly, in this pQCD analysis we use only full seven LHC ATLAS inclusive jet production cross sections data corresponding to radius parameter $R = 0.6$.     

Table~\ref{tab:jdata} shows a summary of full seven LHC ATLAS inclusive jet production cross sections data corresponding to $R = 0.6$ radius parameter.

\begin{table}[h]
\begin{center}
\begin{tabular}{|l|c|c|c|c|c|c|c|c|}
\hline
\hline
 {Experiment} & {NDP} & {parameters} & {inelasticity} & {${\cal L}$} & $\sqrt{s}$ & {Ref.} \\ \hline 
  LHC ATLAS Jet data $0.0$ & 16 & R = $0.6$ & $\mid y \mid < 0.3$ & $37$~pb$^{-1}$ & $7$~TeV & \cite{Aad:2011fc} \\ \hline 
  LHC ATLAS Jet data $0.3$ & 16 & R = $0.6$ & $\mid y \mid < 0.8$ & $37$~pb$^{-1}$ & $7$~TeV & \cite{Aad:2011fc} \\ \hline 
  LHC ATLAS Jet data $0.8$ & 16 & R = $0.6$ & $\mid y \mid < 1.2$ & $37$~pb$^{-1}$ & $7$~TeV & \cite{Aad:2011fc} \\ \hline 
  LHC ATLAS Jet data $1.2$ & 15 & R = $0.6$ & $\mid y \mid < 2.1$ & $37$~pb$^{-1}$ & $7$~TeV & \cite{Aad:2011fc} \\ \hline 
  LHC ATLAS Jet data $2.1$ & 12 & R = $0.6$ & $\mid y \mid < 2.8$ & $37$~pb$^{-1}$ & $7$~TeV & \cite{Aad:2011fc} \\ \hline
  LHC ATLAS Jet data $2.8$ & 9 & R = $0.6$ & $\mid y \mid < 3.6$ & $37$~pb$^{-1}$ & $7$~TeV & \cite{Aad:2011fc} \\ \hline 
  LHC ATLAS Jet data $3.6$ & 6 & R = $0.6$ & $\mid y \mid < 4.4$ & $37$~pb$^{-1}$ & $7$~TeV & \cite{Aad:2011fc} \\ \hline \hline
  \end{tabular}
\vspace{-0.0cm}
\caption{\label{tab:jdata}{A summary of full seven LHC ATLAS inclusive jet production cross sections data.}}
\vspace{-0.4cm}
\end{center}
\end{table}

The inclusive jet double-differential cross-sections $\text{d}^2\sigma_{\rm jet} / dp_T dy$ is defined as a function of jet high transverse momentum $p_T$ in bins of rapidity $y$. The normalized jet production cross sections are defined as a ratio of differential inclusive jet to the differential NC cross section ($\frac{\sigma_{\rm jet}}{\sigma_{\rm NC}}$) in a given $Q^2$ bin , multiplied by the respective bin width $W$ in the case of a double differential measurement as follows:

\begin{eqnarray}\label{jcs}
\frac{\sigma_{\rm jet}}{\sigma_{\rm NC}}\left(Q^2,\, p_T\right) \hspace{1pc} &=& 
\frac{ \text{d}^2\sigma_{\rm jet} / \text{d}Q^2\,\text{d}p_{T} }
     { \text{d}\sigma_{\rm NC}   /  \text{d}Q^2}
     \cdot W(p_T)~.
\end{eqnarray}

The kinematic range of LHC ATLAS inclusive jet production cross sections measurement is $ 20 \leq p_T \leq 430$~GeV corresponding to rapidity $y < \mid4.4\mid$.

Also, we may define the inclusive jet double-differential cross-sections in terms of the invariant cross-sections as follows:

\begin{eqnarray}
   \frac{1}{2\pi p_T} \frac{d^{2}\sigma_{\rm jet}}{dp_T dy} = E \frac{d^{3}\sigma_{\rm jet}}{dp^{3}},
\end{eqnarray}
where $p$ and $E$  refer to the momentum and energy of the jet, respectively.

\section{\label{fit}Fitting and proton PDFs}

As previously mentioned, we parameterized the proton PDFs based on the standard functional form:

\begin{eqnarray}
\label{eq:ff}
xf(x) = Ax^{B}(1 - x)^{C}P_l(x)~,
\end{eqnarray} 

where $P_l(x)$ is an appropriate polynomial function which interpolates between the low and high $x$ regions. Based on the standard functional form the HERAPDF style can be introduced by

\begin{equation}
\label{eq:hps}
 xf(x) = A x^{B} (1-x)^{C} (1 + D x + E x^2)~.
\end{equation}

Using the HERAPDF style, we may parametrize the valence distribution $xu_v$ and $xd_v$ as follows:
\begin{eqnarray}
\label{eq:xuv}
xu_v(x) &=& A_{u_v} x^{B_{u_v}}  (1-x)^{C_{u_v}}\left(1+E_{u_v}x^2 \right)~, \\
\label{eq:xdv}
xd_v(x) &=& A_{d_v} x^{B_{d_v}}  (1-x)^{C_{d_v}}~.
\end{eqnarray}

We set the evolution starting scale at $Q_0^2 = 1.9$~GeV$^2$ and then we evolve the parametrized proton PDFs based on the DGLAP collinear evolution by QCDNUM package as a powerful and very fast QCD evolution program written in FORTRAN$77$~\cite{Botje:2010ay}.

The $x\bar{U}(x)$ and $x\bar{D}(x)$ are $u$-type and $d$-type sea distributions, respectively and they defined as: $x\bar{U}(x) = x\bar{u}(x)$ and $x\bar{D}(x) = x\bar{d}(x) + x\bar{s}(x)$. In the HERAPDF style the parametric form of $u$-type and $d$-type sea distributions are as follows:

\begin{eqnarray}
\label{eq:xubar}
x\bar{U}(x) &= & A_{\bar{U}} x^{B_{\bar{U}}} (1-x)^{C_{\bar{U}}}\left(1+D_{\bar{U}}x\right)~, \\
\label{eq:xdbar}
x\bar{D}(x) &=& A_{\bar{D}} x^{B_{\bar{D}}} (1-x)^{C_{\bar{D}}}~.
\end{eqnarray}

Based on the HERAPDF style, we may write the gluon distribution $xg(x)$ at the first step as follows:

\begin{eqnarray}
\label{eq:xg1}
xg(x) &=& A_g x^{B_g} (1-x)^{C_g}~. 
\end{eqnarray}

However in order to control the $xg(x)$ behavior at low $x$ it is better to extend the Eq.~(\ref{eq:xg1}) to the following form:

\begin{eqnarray}
\label{eq:xg}
xg(x) &=& A_g x^{B_g} (1-x)^{C_g} - A_g' x^{B_g'} (1-x)^{C_g'}. 
\end{eqnarray}

Really, the extra term $A_g' x^{B_g'} (1-x)^{C_g'}$ makes the gluon distribution $xg(x)$ more flexible at low values of $x$. In addition in this pQCD analysis the parameter $C_g'$ is fixed to $C_g'= 25$ to ensure a positive gluon density at large values of Bjorken scaling $x$.

\section{\label{qcdsetup}QCD set-up and minimization}
The proton PDFs are phenomenologically extracted from QCD fits by a measure of the agreement between
experimental data and the QCD theory models. To develop this pQCD analysis we use the following initial QCD set-up and packages:

\begin{itemize}
\item {\bf QCD framework:} We use xFitter version $2.0.0$ as a very powerful an open source QCD fit framework which has been designed to extract proton PDFs and assess the impact of new data~\cite{xFitter,Vafaee:2019hwf,Shokouhi:2018gie,Vafaee:2016jxl,Vafaee:2017jnt,Vafaee:2019yec,Vafaee:2018ehy,Vafaee:2018abd}.

\item {\bf QCD evolution:} We parametrize the proton PDFs based on the HRAPDF style and evolve the parametrized PDFs from starting scale of $Q_0^2 = 1.9$~GeV$^2$ based on the DGLAP collinear evolution equations using QCDNUM package version $17-01/14$~\cite{Botje:2010ay}.

\item {\bf Fast convolution calculations:} We use the hoppet code package version $1.2.0$~\cite{Salam:2008sz} and APPLgrid C++ code package version $1.5.35$ as a fast and flexible approach to reproduce the results of full NLO calculations with input proton PDFs~\cite{Carli:2010rw}.



\item {\bf Minimization:}
As we mentioned, the $\chi^2$-function is a measure of the agreement between experimental data and the QCD theory models. To include of systematic and statistical uncertainties into the $\chi^2$-function definition there are some different approaches, however the correlated systematic uncertainties can be kept separately. We use the following $\chi^2$-function definition to include of correlated and uncorrelated errors:

\begin{equation}\label{eq:chicor}
\chi^2=\sum_{j=1}^{N_{\rm pts}}\left(\frac{D_j+\sum_{k=1}^{N_{\rm corr}}r_k\sigma_{k,j}^{\rm corr}-T_j}{\sigma_j^{\rm uncorr}}\right)+\sum_{k=1}^{N_{\rm corr}} r_k^2\;,
\end{equation}
where $D_j$ is $j$th data point, $r_k$ is the size of the shifts for each source of systematic uncertainty, $\sigma^{\rm corr}_{k,j}$ is the correlated errors, $T_j$ is the theory prediction  and $\sigma^{\rm uncorr}_j$ is the uncorrelated errors. The minimization procedure of this pQCD N3LO analysis has been done based on the standard MINUIT-minimization program~\cite{James:1975dr} with $14$ unknown fit parameters in HERAPDF style functional form.
\end{itemize}

\section{\label{res}Results}
Based on our QCD set-up and minimization procedure, we determine the numerical values of $14$ unknown fit parameters as illustrated in Table~\ref{tab:par}. 

\begin{table}[h]
\begin{center}
\begin{tabular}{|l|c|c|c|c|}
\hline
\hline
 \multicolumn{5}{|c|}{ Numerical values of $14$ fit parameters}    \\ \hline
 {PARAMETER} & {HNLO} & {HANLO} & {HAN2LO} & {HAN3LO} \\ \hline
  ${B_{u_v}}$ & $0.730 \pm 0.042$& $0.685 \pm 0.030$& $0.797 \pm 0.028$& $0.774 \pm 0.029$ \\ \hline 
  ${C_{u_v}}$ & $4.827 \pm 0.083$& $4.827 \pm 0.076$& $4.768 \pm 0.081$& $4.705 \pm 0.079$ \\ \hline
  $E_{u_v}$ &  $13.1 \pm 2.0$& $14.9 \pm 1.8$& $10.0 \pm 1.3$& $10.4 \pm 1.3$ \\ \hline 
  ${B_{d_v}}$ & $0.83 \pm 0.13$& $0.833 \pm 0.084$& $1.003 \pm 0.091$& $0.960 \pm 0.083$ \\ \hline
  $C_{d_v}$ & $4.21 \pm 0.40$& $4.29 \pm 0.35$& $4.88 \pm 0.38$& $4.69 \pm 0.33$ \\ \hline 
  $C_{\bar{U}}$ & $8.90 \pm 0.76$& $7.61 \pm 0.71$& $6.6 \pm 1.3$& $7.09 \pm 0.97$ \\ \hline 
  $D_{\bar{U}}$ & $17.6 \pm 3.1$& $11.2 \pm 2.0$& $2.3 \pm 1.8$& $6.2 \pm 1.7$  \\ \hline
  $A_{\bar{D}}$ & $0.156 \pm 0.010$& $0.1697 \pm 0.0091$& $0.2448 \pm 0.0098$& $0.1870 \pm 0.0072$ \\ \hline
  $B_{\bar{D}}$ & $-0.1760 \pm 0.0076$& $-0.1593 \pm 0.0065$& $-0.1290 \pm 0.0049$& $-0.1632 \pm 0.0046$ \\ \hline 
  $C_{\bar{D}}$ & $4.0 \pm 1.9$& $6.2 \pm 1.1$& $8.9 \pm 2.0$& $6.8 \pm 1.1$ \\ \hline
  $B_g$ & $-0.069 \pm 0.075$& $-0.051 \pm 0.066$& $-0.065 \pm 0.047$& $-0.150 \pm 0.039$ \\ \hline 
  $C_g$ & $12.3 \pm 1.0$& $7.81 \pm 0.62$& $6.43 \pm 0.54$& $7.49 \pm 0.51$  \\ \hline 
  $A_g'$ & $2.89 \pm 0.49$& $0.49 \pm 0.13$& $0.217 \pm 0.058$& $1.167 \pm 0.093$  \\ \hline 
  ${B_g'}$ & $-0.142 \pm 0.062$& $-0.277 \pm 0.033$& $-0.360 \pm 0.029$& $-0.257 \pm 0.024$ \\ \hline \hline
    \end{tabular}
\vspace{-0.0cm}
\caption{\label{tab:par}{ {The numerical values of $14$ fit parameters corresponding to four different HNLO, HANLO, HAN2LO and HAN3LO pQCD analysis.}}}
\vspace{-0.4cm}
\end{center}
\end{table}

Table~(\ref{tab:fdata}) shows the experimental data, correlated ${\chi^2}$ and partial $\chi^2$ per degree of freedom (dof) for each experiment  corresponding to four different HNLO, HANLO, HAN2LO and HAN3LO pQCD analysis.

\begin{table}[h]
\begin{center}
\begin{tabular}{|l|c|c|c|c|}
\hline
\hline
 \multicolumn{5}{|c|}{ Global fit based on four different HNLO, HANLO, HAN2LO and HAN3LO analysis }    \\ \hline
 {EXPERIMENT} & {HNLO} & {HANLO} & {HAN2LO} & {HAN3LO} \\ \hline 
  HERA I+II CC $e^{+}p$ \cite{Abramowicz:2015mha} & 44 / 39& 49 / 39& 47 / 39& 45 / 39 \\ \hline 
  HERA I+II CC $e^{-}p$ \cite{Abramowicz:2015mha} & 50 / 42& 49 / 42& 51 / 42& 49 / 42 \\ \hline 
  HERA I+II NC $e^{-}p$ \cite{Abramowicz:2015mha} & 221 / 159& 221 / 159& 219 / 159& 216 / 159 \\ \hline
  HERA I+II NC $e^{+}p$ 460 \cite{Abramowicz:2015mha} & 210 / 204& 211 / 204& 212 / 204& 213 / 204 \\ \hline
  HERA I+II NC $e^{+}p$ 575 \cite{Abramowicz:2015mha} & 212 / 254& 217 / 254& 213 / 254& 215 / 254 \\ \hline
  HERA I+II NC $e^{+}p$ 820 \cite{Abramowicz:2015mha} & 65 / 70& 67 / 70& 64 / 70& 67 / 70 \\ \hline
  HERA I+II NC $e^{+}p$ 920 \cite{Abramowicz:2015mha} & 418 / 377& 439 / 377& 427 / 377& 425 / 377  \\ \hline
LHC ATLAS Jet data $0.0$ \cite{Aad:2011fc} &  - & 15 / 16& 15 / 16& 15 / 16  \\ \hline
LHC ATLAS Jet data $0.3$ \cite{Aad:2011fc} &  - & 9.3 / 16& 6.5 / 16& 8.8 / 16  \\ \hline
LHC ATLAS Jet data $0.8$ \cite{Aad:2011fc} &  - & 9.4 / 16& 6.8 / 16& 8.1 / 16  \\ \hline
LHC ATLAS Jet data $1.2$ \cite{Aad:2011fc} &  - & 8.0 / 15& 7.4 / 15& 7.7 / 15  \\ \hline
LHC ATLAS Jet data $2.1$ \cite{Aad:2011fc} &  - & 6.8 / 12& 6.9 / 12& 6.8 / 12  \\ \hline
LHC ATLAS Jet data $2.8$ \cite{Aad:2011fc} &  - & 1.8 / 9& 1.7 / 9& 1.9 / 9  \\ \hline
LHC ATLAS Jet data $3.6$ \cite{Aad:2011fc} &  - & 0.85 / 6& 0.80 / 6& 0.78 / 6  \\ \hline 
 {Correlated ${\chi^2}$} & 111& 111& 119& 117 \\ \hline
{${\frac{{\chi^2}_{Total}}{dof}}$} & ${\frac{1331}{1131}}$  & ${\frac{1415}{1221}}$ &  ${\frac{1397}{1221}}$ & ${\frac{1396}{1221}}$ \\ \hline \hline
  \end{tabular}
\vspace{-0.0cm}
\caption{\label{tab:fdata}{Experimental data, correlated ${\chi^2}$ and partial $\chi^2$ per degree of freedom (dof) for each experiment  corresponding to four different HNLO, HANLO, HAN2LO and HAN3LO pQCD analysis.}}
\vspace{-0.4cm}
\end{center}
\end{table}

As we mentioned, our central proton PDFs are extracted in HNLO QCD analysis by fitting with HERA I and II combined data. In order to estimate the quality of the fit due to inclusion of the LHC ATLAS jet cross sections data, we compare the relative improvement of the quality of the fit corresponding to three different QCD HANLO, HAN2LO and HAN3LO QCD analysis with HNLO analysis as our main QCD fit for extracting the central proton PDFs.
    
Table~(\ref{tab:fq}) shows a comparison between {$\frac{{\chi^2}_{Total}}{dof}$}, the QCD quality of the fit and relative improvement in the quality of the fit corresponding to four different HNLO, HANLO, HAN2LO and HAN3LO pQCD analysis.

\begin{table}[h]
\begin{center}
\begin{tabular}{|l|c|c|c|c|}
\hline
\hline
 {QCD analysis} & {{$\frac{{\chi^2}_{Total}}{dof}$}} & {fit-quality} & {relative improvement}  \\ \hline 
HANLO & ${\frac{1415}{1221}}$ & $1.158$ & ${\frac{\vert\chi^2_{\rm HANLO}-\chi^2_{\rm HNLO}\vert}{\chi^2_{\rm HNLO}} = \frac{\vert1.158 - 1.176\vert}{1.176}} \sim 1.53$~\% \\ \hline
HAN2LO & ${\frac{1397}{1221}}$ & $1.144$ & ${\frac{\vert\chi^2_{\rm HAN2LO}-\chi^2_{\rm HNLO}\vert}{\chi^2_{\rm HNLO}} = \frac{\vert1.144 - 1.176\vert}{1.176}} \sim 2.72$~\% \\ \hline
HN3LO & ${\frac{1396}{1221}}$ & $1.143$ & ${\frac{\vert\chi^2_{\rm HAN3LO}-\chi^2_{\rm HNLO}\vert}{\chi^2_{\rm HNLO}} = \frac{\vert1.143 - 1.176\vert}{1.176}} \sim 2.80$~\% \\ \hline \hline
  \end{tabular}
\vspace{-0.0cm}
\caption{\label{tab:fq}{Comparison the relative improvement of the quality of the fit corresponding to three different QCD HANLO, HAN2LO and HAN3LO pQCD analysis with HNLO analysis as our main QCD fit for extracting the central proton PDFs.}}
\vspace{-0.4cm}
\end{center}
\end{table}

\begin{figure*}
\includegraphics[width=0.49\textwidth]{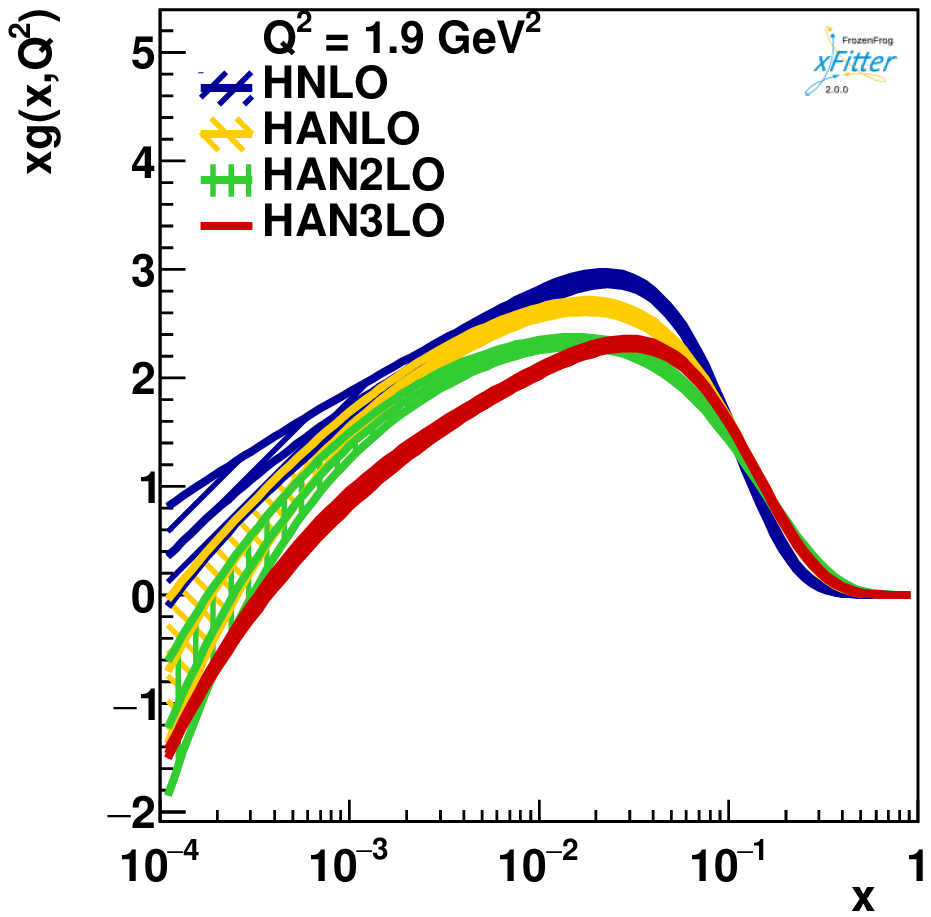}
\includegraphics[width=0.49\textwidth]{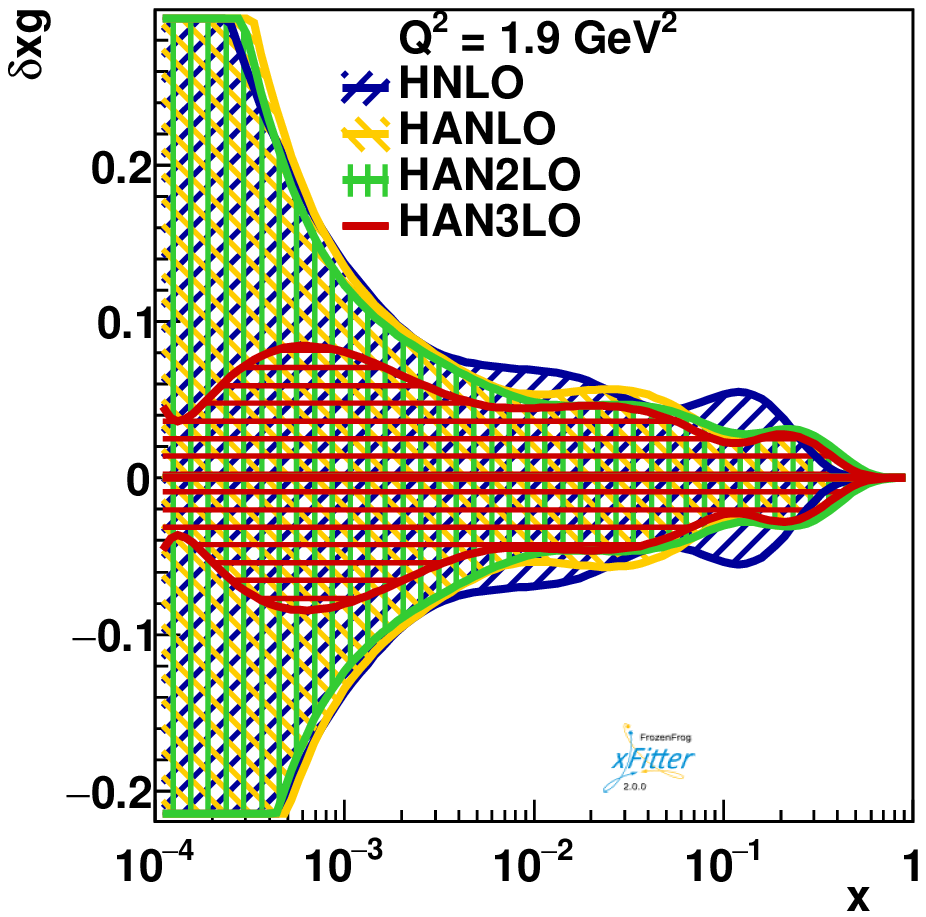}

\includegraphics[width=0.49\textwidth]{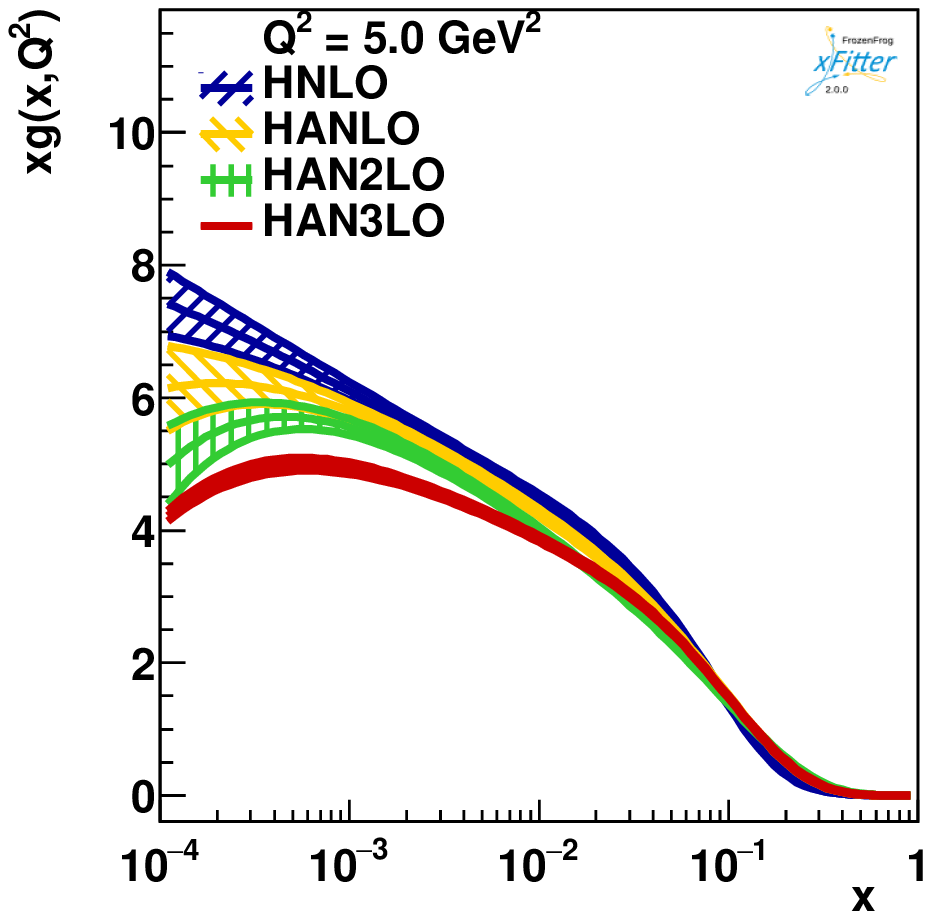}
\includegraphics[width=0.49\textwidth]{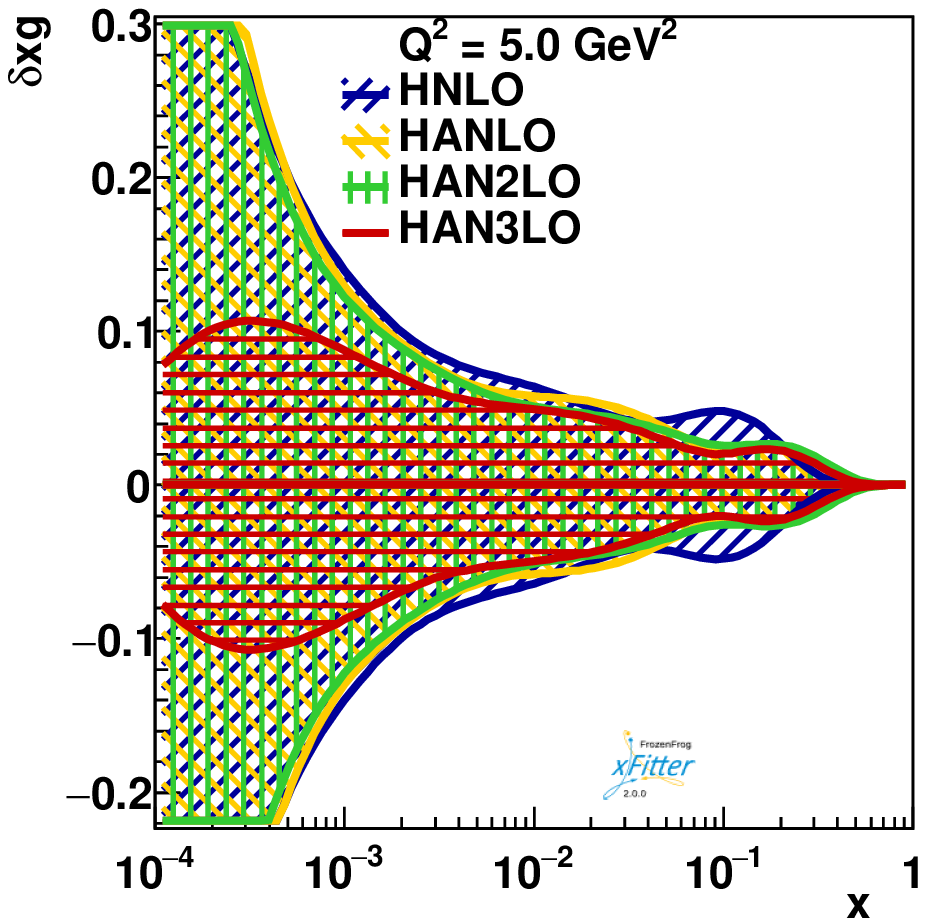}

\includegraphics[width=0.49\textwidth]{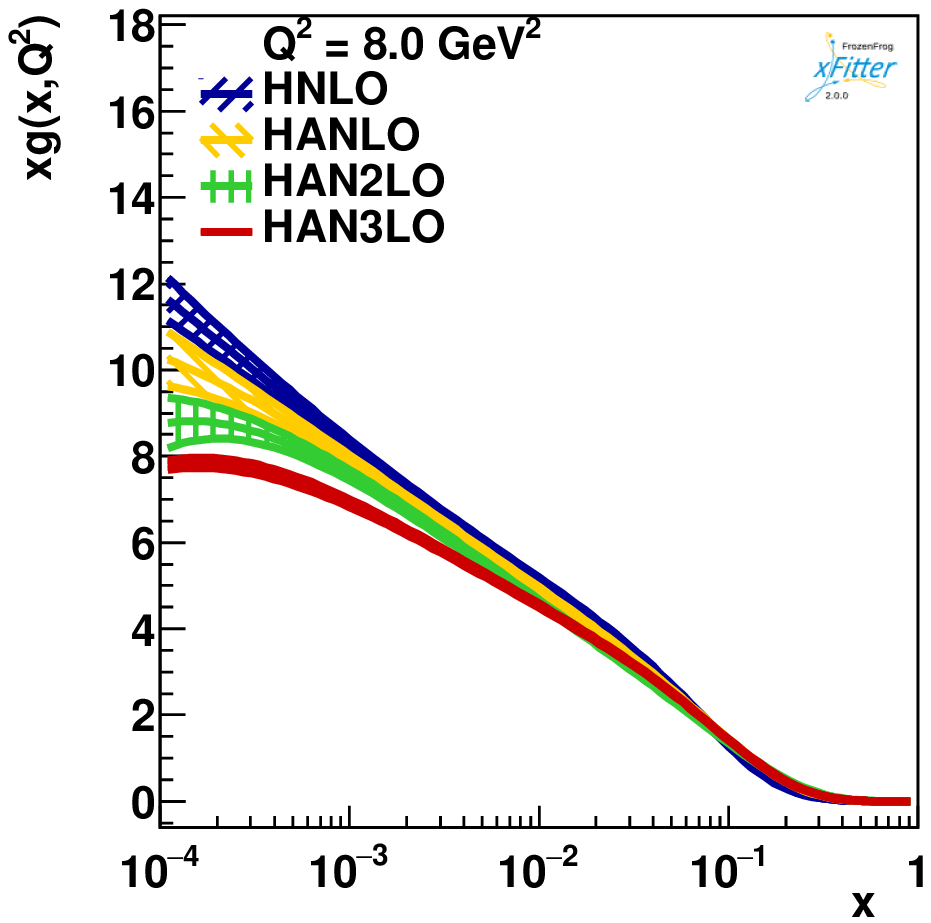}
\includegraphics[width=0.49\textwidth]{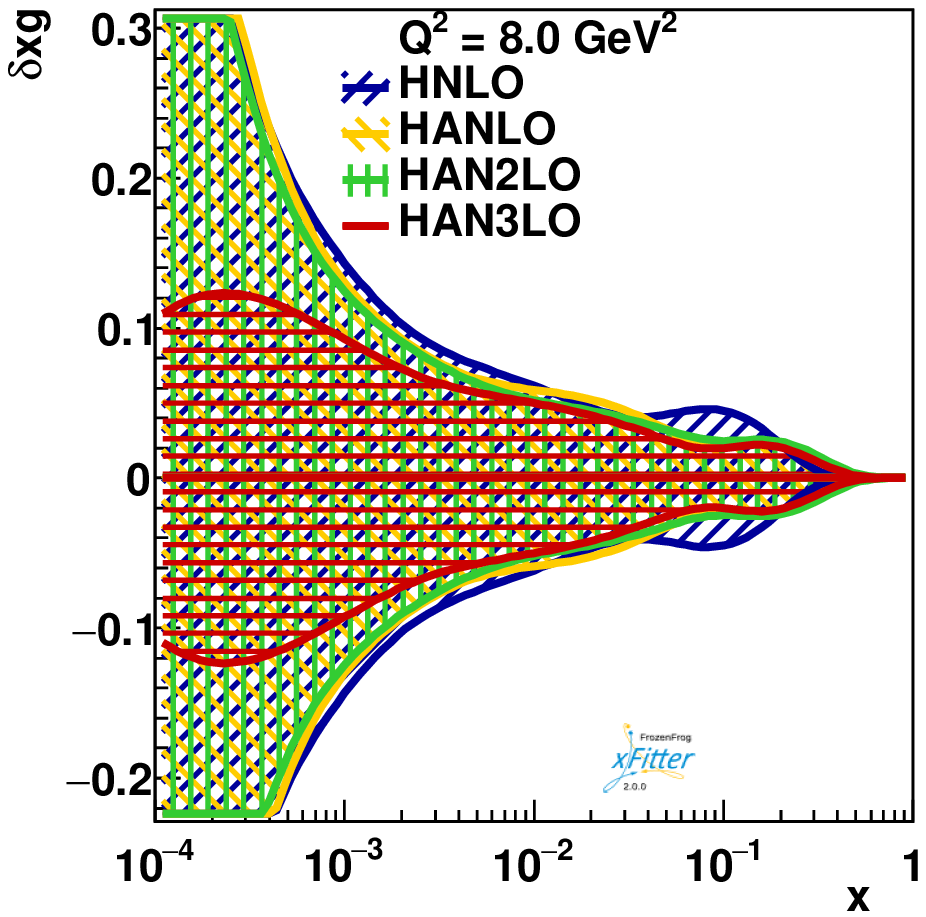}
\caption{Impact of inclusion of the LHC ATLAS jet production cross sections data on the HERA I and II combined data for gluon and its ratio distributions corresponding to four different HNLO, HANLO, HAN2LO and HAN3LO analysis}.
\label{fig:1}
\end{figure*}

Fig.~(\ref{fig:1}) shows the impact of inclusion of the LHC ATLAS jet production cross sections data on the HERA run I and II combined data as our central proton PDFs for gluon and its ratio distributions corresponding to four different HNLO, HANLO, HAN2LO and HAN3LO analysis at the starting scale of $Q_0^2 = 1.9$~GeV$^2$ and $Q^2 = 5$ and  $8$~GeV$^2$.

\begin{figure*}
\includegraphics[width=0.49\textwidth]{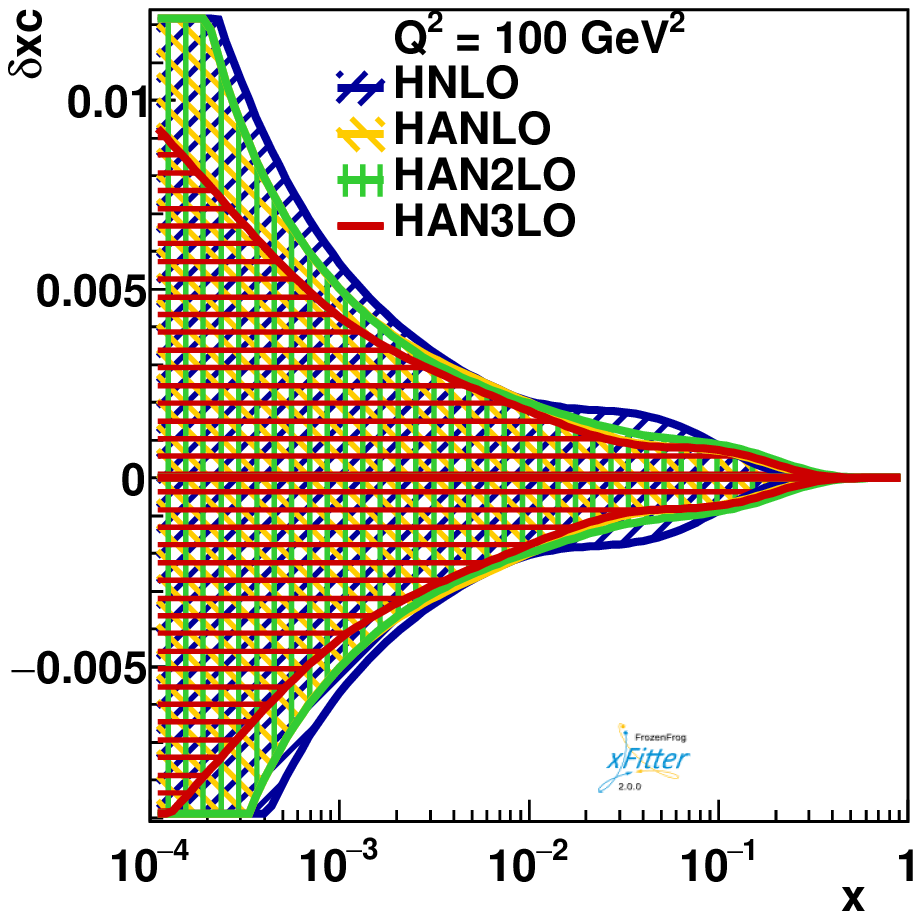}
\includegraphics[width=0.49\textwidth]{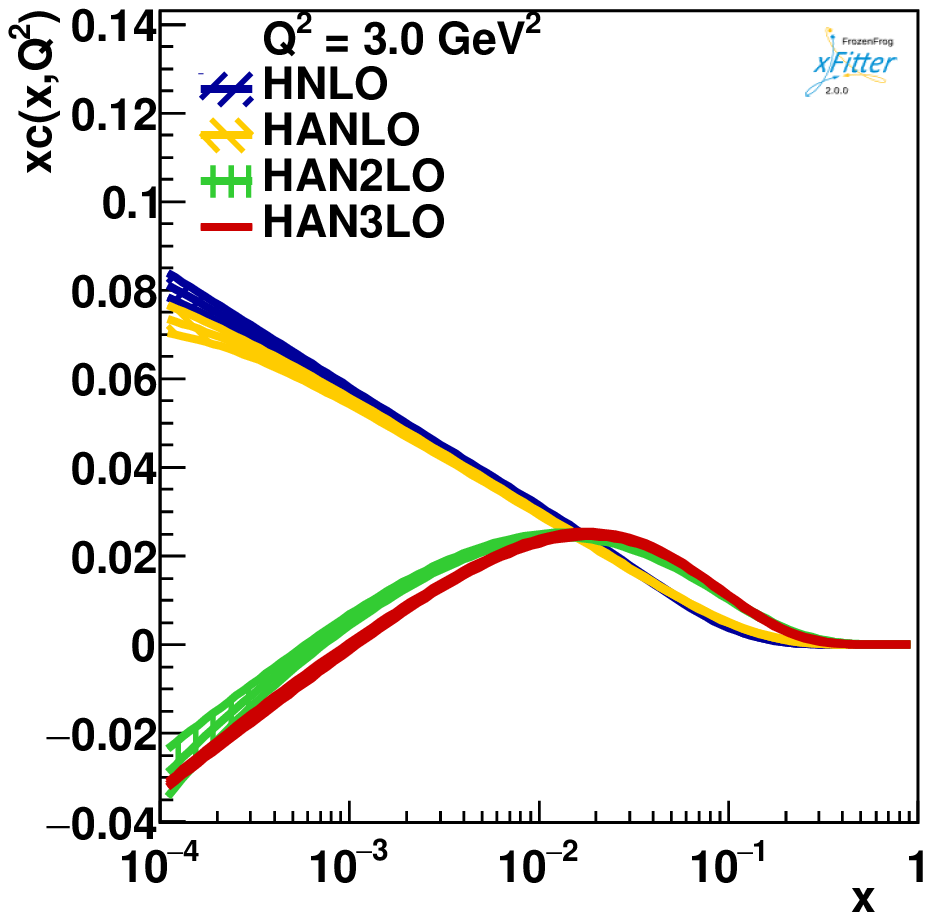}

\includegraphics[width=0.49\textwidth]{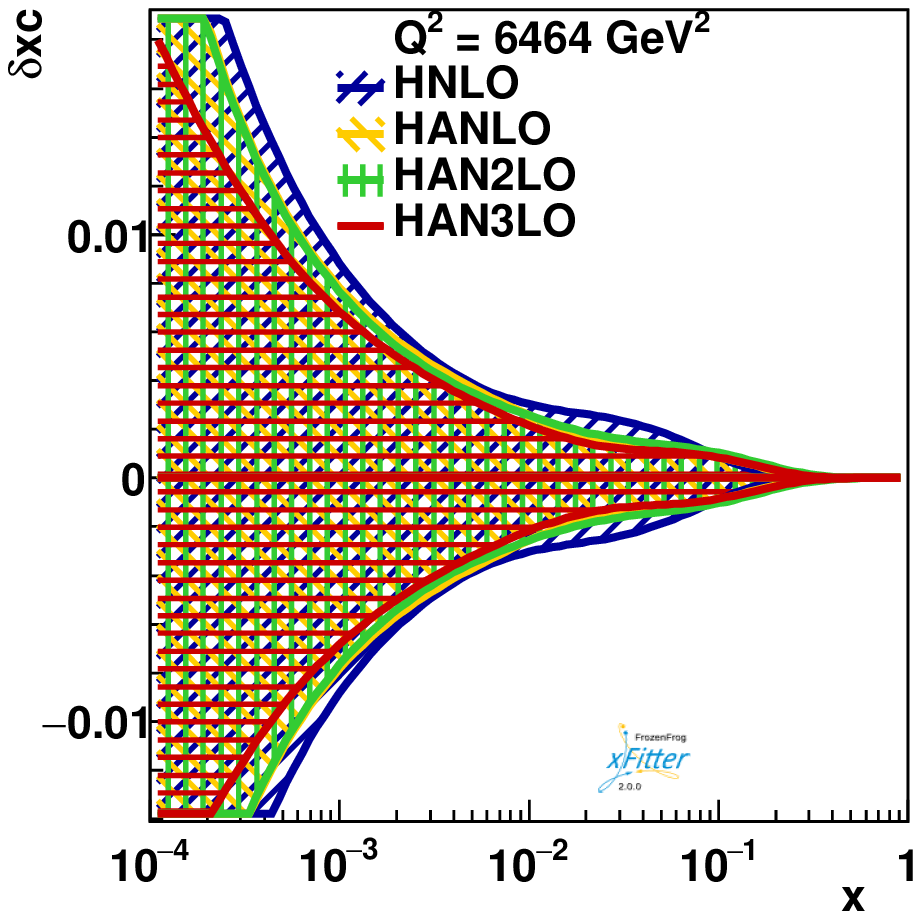}
\includegraphics[width=0.49\textwidth]{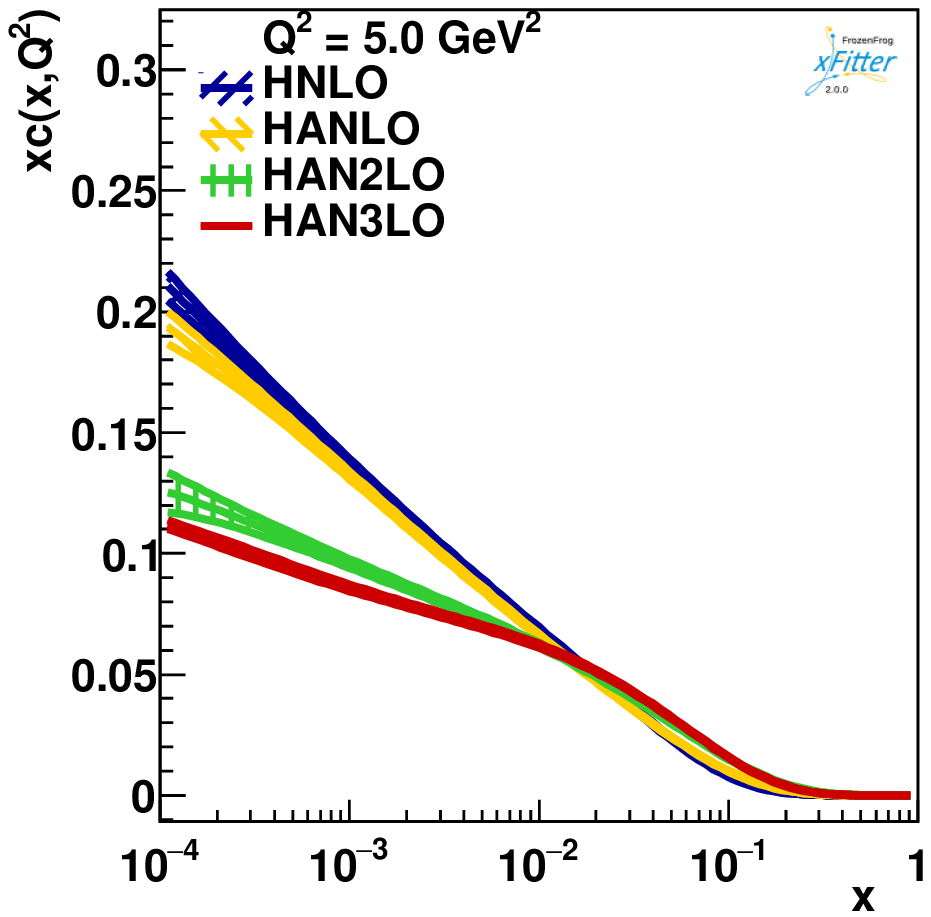}

\includegraphics[width=0.49\textwidth]{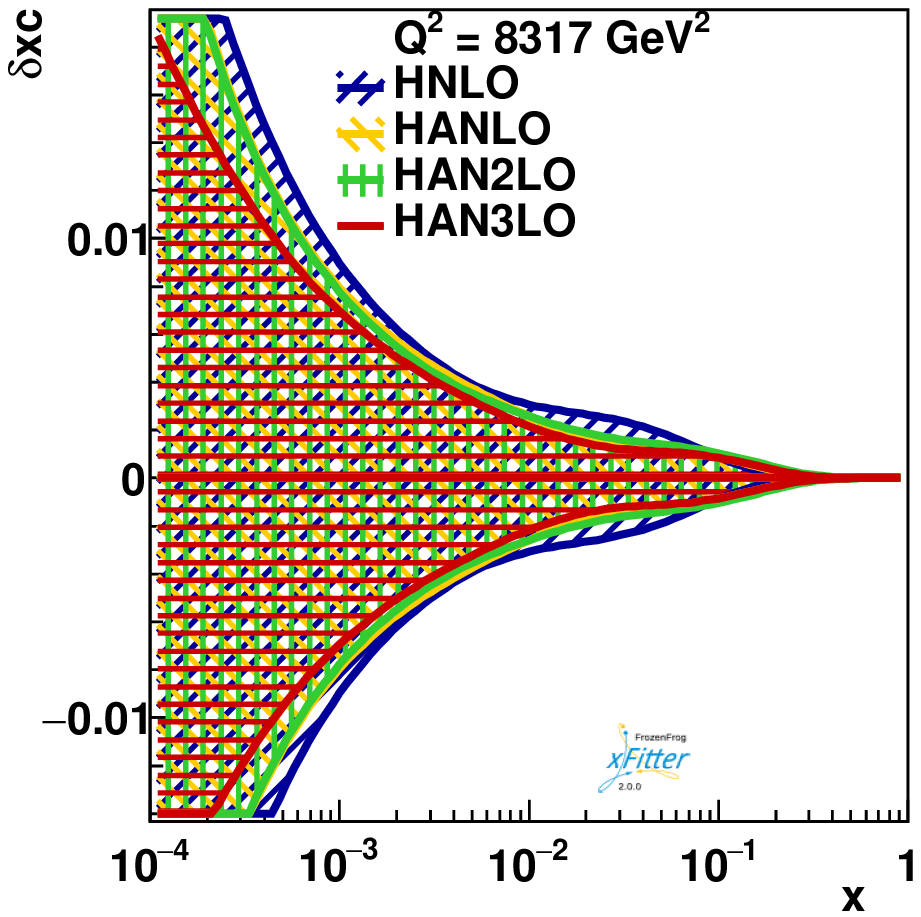}
\includegraphics[width=0.49\textwidth]{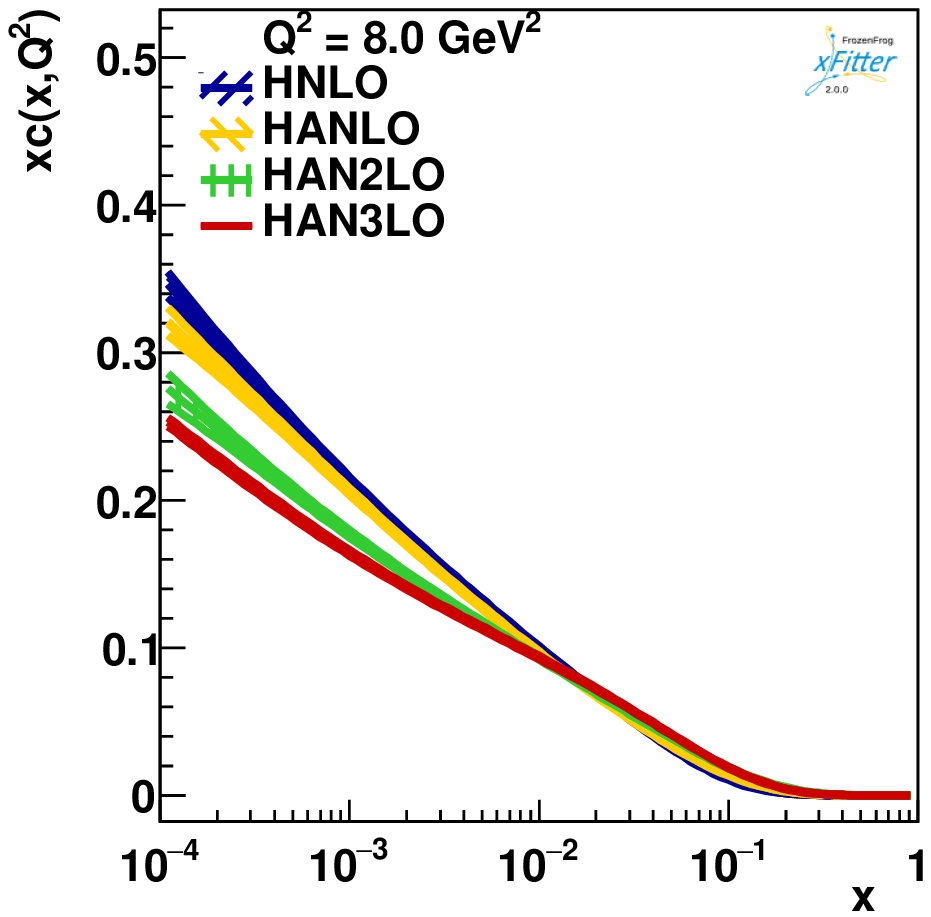}
\caption{Impact of inclusion of the LHC ATLAS jet production cross sections data on the HERA run I and II combined data as our central proton PDFs for $xc(x,Q^2)$ and its ratio distributions corresponding to four different HNLO, HANLO, HAN2LO and HAN3LO analysis}.
\label{fig:2}
\end{figure*}

In Fig.~(\ref{fig:2}) we illustrate the impact of inclusion of the LHC ATLAS jet production cross sections data on the HERA run I and II combined data as our central proton PDFs for $xc(x,Q^2)$ and its ratio distributions corresponding to four different HNLO, HANLO, HAN2LO and HAN3LO analysis at $Q^2 = 3, 5, 8, 100, 6464$ and $8317$~GeV$^2$.  

\section{\label{sum}Summery and discussion}
\begin{itemize}
\item From point of view of quark-parton model, we may obtain much more direct information about quark and gluon content of the proton by global fit of available experimental data from different NC and CC reactions at the DIS level experiments.

\item Phenomenologically, we define the proton as a whole by extracting the proton PDFs based on the HERA run I and II combined data (HNLO QCD analysis) and then we take them into account as the central proton PDFs in our QCD analysis.

\item We add the full seven set of the LHC ATLAS jet cross sections data at $ \sqrt{s} = 7$~TeV on the proton central PDFs to investigate the role and influence of these data on the proton PDFs at the NLO, N2LO and N3LO corrections.   

\item We find that inclusion of the the LHC ATLAS jet cross sections data on the central proton PDFs reduces the error band of proton PDFs, particularly reduces dramatically the uncertainties of gluon $xg(x,Q^2)$ and charm $xc(x,Q^2)$ distributions (HANLO, HAN2LO and HAN3LO QCD analysis).

\item We find that adding the LHC ATLAS jet cross sections data on the central proton PDFs improves the quality of the fit up to $\sim 1.53$~\%, $\sim 2.72$~\% and $\sim 2.80$~\% corresponding to NLO, N2LO and N3LO QCD analysis, respectively.

\item According to the Figs.~(\ref{fig:1})~-~(\ref{fig:2}), the best improvements in the uncertainty of the proton PDFs are corresponding to the HAN3LO QCD analysis.

\item According to the Table~(\ref{tab:fq}), the best improvement in the quality of the fit is up to $\sim 2.80$~\% corresponding to HAN3LO QCD analysis.     

\item Standard LHAPDF files corresponding to HNLO, HANLO, HAN2LO and HAN3LO QCD analysis are available      for the fast QCD analysis based on the profiling and reweighting techniques and can be obtain from authors via e-mail.

\end{itemize}

\section{Acknowledgments}
We are grateful to Prof. F. Olness from SMU (Southern Methodist University) for developing invaluable heavy flavor schemes, particularly FONLL class schemes as implemented in the xFitter package. We are also grateful to Dr. Francesco Giuli, Dr. Ivan Novikov, Dr. Oleksandr Zenaiev and Dr. Sasha Glazov from xFitter developer group for very invaluable comments and suggestions about installation of ROOT version $6.16/00$ and xFitter version $2.0.1$ on SL7 platform. We would like to appreciate Mrs. Malihe Shokouhi for spending time and careful reading the draft version of this manuscript. This work is related to the ``Special Support Program for the Promotion of Scientific Authority'' in Ferdowsi University of Mashhad.    


\end{document}